\pdfoutput=1

\documentclass[a4paper]{article}

\providecommand{\keywords}[1]{\textbf{Keywords: } #1}

\usepackage[T1]{fontenc}
\usepackage{lmodern}
\usepackage{amsfonts}

\usepackage{amsmath}
\DeclareMathOperator*{\argmax}{arg\,max}

\usepackage{enumitem}

\newcommand{\appendixnumberline}[1]{Appendix\space}
\let\oldappendix\appendix
\makeatletter
\renewcommand{\appendix}{%
  \addtocontents{toc}{\let\protect\numberline\protect\appendixnumberline}%
  \renewcommand{\@seccntformat}[1]{Appendix~\csname the##1\endcsname\quad}%
  \oldappendix
}
\makeatother

\usepackage[symbol]{footmisc}

\usepackage{cite}

\usepackage{microtype}
\DisableLigatures[f]{encoding = *, family = * }

\usepackage[english]{babel}
\usepackage[utf8x]{inputenc}
\usepackage[T1]{fontenc}

\usepackage[a4paper,top=3cm,bottom=2cm,left=3cm,right=3cm,marginparwidth=1.75cm]{geometry}

\usepackage{amsmath}
\usepackage{graphicx}
\graphicspath{ {graphics/} }
\usepackage[colorinlistoftodos]{todonotes}
\usepackage[colorlinks=true, allcolors=blue]{hyperref}

\begin{document}
\large{This is an Original Manuscript of an article published by Taylor \& Francis in \textit{Brain-Computer Interfaces} on December 6, 2018, available online:\newline \url{http://www.tandfonline.com/10.1080/2326263X.2018.1552357}}
\pagenumbering{gobble}
\clearpage
\setcounter{page}{1}
\pagenumbering{arabic}

\begin{flushleft}
{\Large
\textbf\newline{An Analysis of the Accuracy of the P300 BCI}
}
\newline
\\
Nitzan S. Artzi\textsuperscript{1},
Oren Shriki\textsuperscript{1,2,3,*}
\\
\bigskip
\small{\bf{1} Dept. of Cognitive and Brain Sciences, Ben-Gurion University of the Negev, Israel

\bf{2} Dept. of Computer Science, Ben-Gurion University of the Negev, Israel

\bf{3} The Inter-Faculty School for Brain Sciences, Zlotowski Center for Neuroscience, Ben-Gurion University of the Negev, Israel
\\
\bigskip
* shrikio@bgu.ac.il}

\end{flushleft}

\begin{abstract}
The P300 Brain-Computer Interface (BCI) is a well-established communication channel for severely disabled people. The P300 event-related potential is mostly characterized by its amplitude or its area, which correlate with the spelling accuracy of the P300 speller. Here, we introduce a novel approach for estimating the efficiency of this BCI by considering the P300 signal-to-noise ratio (SNR), a parameter that estimates the spatial and temporal noise levels and has a significantly stronger correlation with spelling accuracy. Furthermore, we suggest a Gaussian noise model, which utilizes the P300 event-related potential SNR to predict spelling accuracy under various conditions for LDA-based classification. We demonstrate the utility of this analysis using real data and discuss its potential applications, such as speeding up the process of electrode selection.
\end{abstract}


\section{Introduction} 

One of the most practical BCI paradigms is the so-called P300 BCI, which enjoys the benefits of a relatively short training session and a high information transfer rate \cite{Farwell1988TalkingPotentials,Wolpaw2002Brain-computerControl.,Rak2012Brain-computerPaper}. This BCI paradigm utilizes the P300 event-related potential (ERP), which normally exhibits a positive change in EEG measured voltage and a latency of 250-500 milliseconds. The P300 intensity is more significant for stimuli which are less common (the "Oddball Paradigm"), and typically stronger in the parietal lobe. The exact shape of the P300 wave may vary dramatically from subject to subject \cite{Wolpaw2012Brain-computerPractice}.

\begin{figure}
\centering
\includegraphics[width=0.6\textwidth]{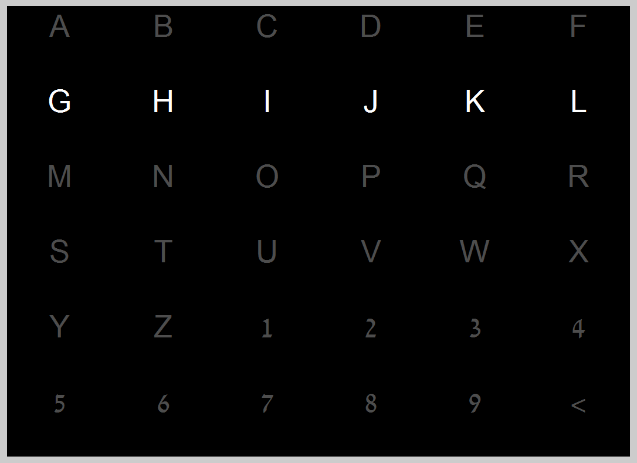}
\caption{\label{fig:1}The character matrix with the second row intensified, as it was presented to the subjects in our system.}
\end{figure}

In the common approach to P300 BCI – often termed the \textit{P300 Speller} – a list of states, organized as a matrix, is displayed to the subject, as shown on Figure \ref{fig:1}.  A row or column is selected randomly, and its intensity is increased for a short period of time ("flash duration"), a process which is repeated with a constant interval between flashes ("inter-stimulus interval", or ISI), until all rows and columns have been flashed. The subject is then asked to count the number of flashes of the intended target stimulus. Repeating the above several times for each row or column and then averaging the signal recorded for each row and column leads to a significant reduction of the noise, and allows detecting the P300 ERP, which theoretically can be found in (exactly) one row and one column – and their intersection is the selected stimulus.

\subsection{The P300 BCI detection mechanism}

Since the P300 response to every stimulus has a characteristic profile and the information lies solely in the time this response is generated, the following detection scheme is usually used \cite{Farwell1988TalkingPotentials,Fazel-Rezai2011P300-basedDesign}: for the i'th flash, we consider \(E\) electrodes and a time window of \(T\) samples and regard the concatenation of the electrodes signal as an \(E\cdot T\)-dimensional sample, \(x_i\in\mathbb{R}^{E\cdot T}\). The training phase and the detection phase are essentially different: first, we train a model from the single-trial signals, in which a classification problem is defined by labeling \(y_i=1\) for trials in which the target character is contained in the flashed row or column, and \(y_i=0\) otherwise.\footnote[2]{Note that since we only use the classification score \(s\) and not the class \(\argmax{s}\), this is not a classification problem per-se, and in fact can be regarded as a regression problem. In order to retain the usual notation and to be able to compare sample-wise and symbol-wise accuracy, we will use a classification notation.} In order to detect a target symbol, the mean signal for each stimulus is calculated, and then the classifier is evaluated on each averaged signal. The selected symbol is then the intersection of the row and the column for which the classifier score is the highest.

While any classifier could be utilized in this scheme, typically an LDA classifier is employed \cite{Lotte2007AInterfaces}. In LDA, the classification is achieved by multiplying the sample \(x\) with a weight vector \(w\) to obtain \(s=w^T x\), where \(w\) is defined according to
\[w=\hat{\Sigma}^{-1}\cdot\left(\hat{\mu}_1-\hat{\mu}_0\right)\]
where \(\hat{\mu}_1\) and \(\hat{\mu}_0\) are estimates for the mean of the inputs corresponding to \(y=1\) and \(y=0\) accordingly, and \(\hat{\Sigma}\) is an estimate for the within-class covariance matrix. To obtain a decision, the dot product \(s=w^T x\) is compared to a threshold, which is calculated separately \cite{Hastie2001TheLearning}.

\subsection{P300 BCI quality measures}

The main measure of quality for the P300 speller is its accuracy: that is, the ratio between the number of correctly detected symbols to the number of attempted ones, which we will denote the \textbf{symbol selection accuracy}. Note that this quantity is \textit{not} the accuracy of the single-trial classification problem described above (the “classification accuracy”): we consider every symbol rather than every stimulus. The specific method of estimating this ratio may vary from one study to another, and can be calculated either online or offline. Other quality measures include the \textit{Information Transfer Rate} and the \textit{Practical Bit Rate} \cite{Townsend2010AColumns.,Wolpaw1998EEG-basedVerification}, both of which attempt to represent the channel capacity of the BCI. While the information transfer rate and the practical bit rate hold a different meaning, they are both uniquely determined by accuracy, so here we will address accuracy alone.

\subsection*{}

In this article, we show that P300 BCI symbol selection accuracy can be estimated from the signal-to-noise ratio of the single-trial signal. We do this by modeling the P300 signal generation as a simple additive Gaussian noise mechanism, and in practice we show a monotonic relation between signal to noise ratio and accuracy. We demonstrate the above experimentally, and discuss the possible applications.

The rest of the article is structured as follows: in section \ref{section:Estimating} we describe the model and work out its accuracy outcome; in section \ref{section:Application} we present the experimental results and advantages, and in section \ref{section:Discussion} we provide a brief conclusion and discussion.

\section{Estimating the accuracy of the P300 BCI} 
\label{section:Estimating}

We propose the following probabilistic model:

\begin{enumerate}
\item 	The base P300 signal is generated deterministically, so it should have been detected at the EEG as \(\mu_1\) whenever it is generated. When no P300 signal is generated, one would detect a baseline signal of \(\mu_0\).
\item	The actual sampled signal is the P300 wave plus some noise, either 
\(x=\mu_1+z\) or \(x=\mu_0+z\). The noise is distributed normally with mean zero and covariance matrix \(\Sigma\). All the noise vectors are sampled i.i.d, and are independent of the presence of the P300 signal.
\item	The detection mechanism is as described in section 1.1, with an LDA classifier. Although the weight vector of the classifier is determined by a contingent training set and is therefore random, we assume for the sake of simplicity that it is indeed the Bayesian optimum of \(w=\Sigma^{-1}\cdot\left(\mu_1-\mu_0\right)\).
\end{enumerate}

As we derive in detail in Appendix \ref{app:A}, the accuracy is determined according to the \textit{accuracy function}, \(H_N\left(x\right)\), defined as
\[H_N\left(x\right)\stackrel{\text{def}}{=}\intop_{-\infty}^\infty \phi\left(z-x\right)\cdot\Phi^{N-1}\left(z\right) dz\]
where \(\phi\) and \(\Phi\) are the PDF and the CDF of a standard normal distribution, respectively. This function represents the selection accuracy of one out of \(N\) alternatives, given an effective SNR of \(x\). The estimated symbol selection accuracy \(Acc\) will be
\[\begin{split}
Acc &=H_{N_{row}}\left(\sqrt{n}\cdot\gamma\right)\cdot H_{N_{col}}\left(\sqrt{n}\cdot\gamma\right)= \\
= &\left(\intop_{-\infty}^\infty \phi\left(z-\sqrt{n}\cdot\gamma\right)\cdot\Phi^{N_{row}-1}\left(z\right) dz\right)\cdot \\
\cdot &\left(\intop_{-\infty}^\infty \phi\left(z-\sqrt{n}\cdot\gamma\right)\cdot\Phi^{N_{col}-1}\left(z\right) dz\right)
\end{split}\]
where \(\gamma=\sqrt{\left(\mu_1-\mu_0\right)^T\Sigma^{-1}\left(\mu_1-\mu_0\right)}\) is the single-trial signal-to-noise ratio.

\begin{figure}
\centering
\includegraphics[width=0.8\textwidth]{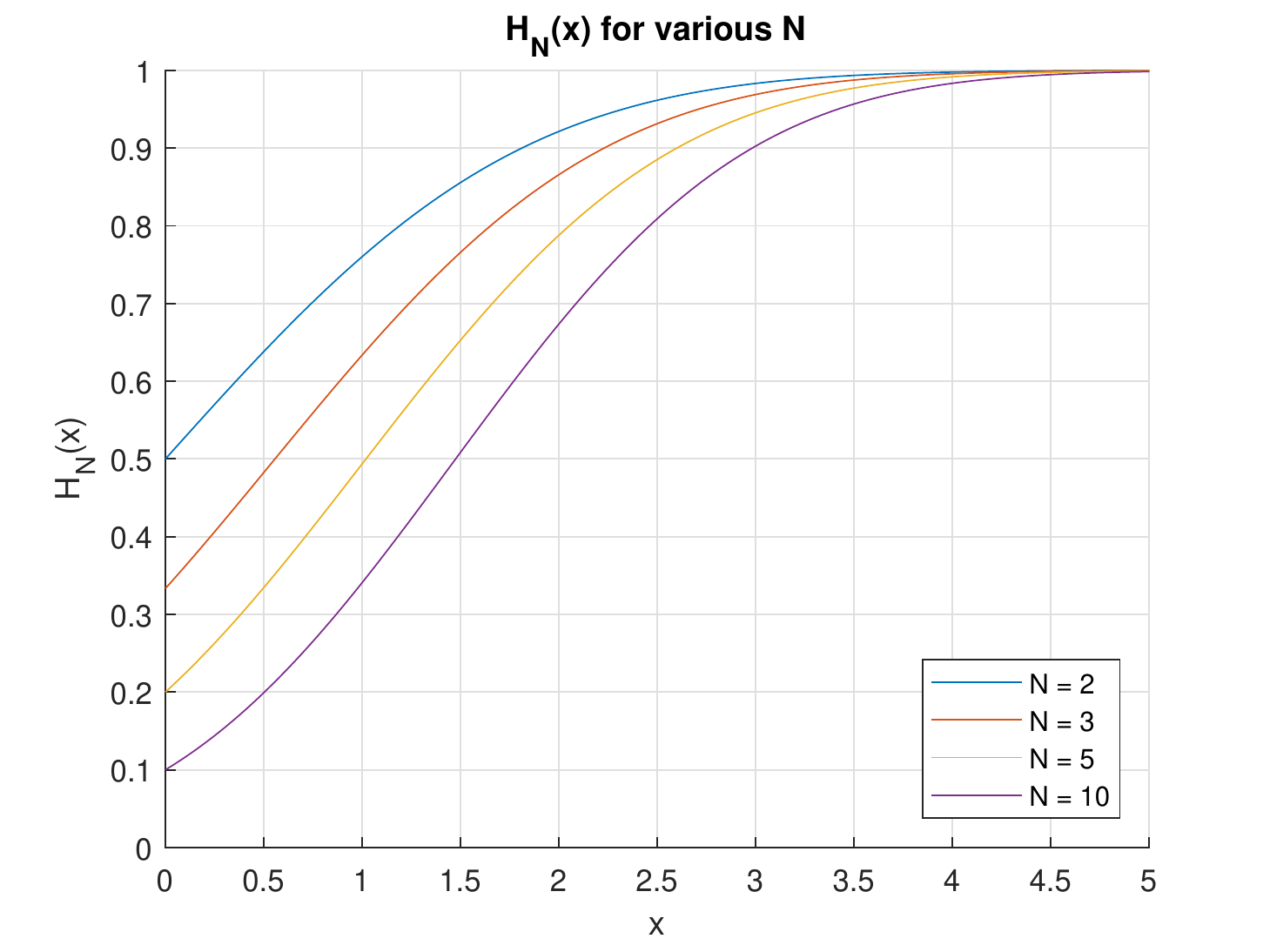}
\caption{\label{fig:2}A numerical calculation of the accuracy function \(H_N\left(x\right)\), depicted for various \(N\)s. Note that the baseline level (\(x=0\)) is indeed the chance probability of selecting one out of \(N\).}
\end{figure}

We can calculate \(H_N\left(x\right)\) numerically, and then infer the selection accuracy and its dependence on the signal-to-noise ratio and the number of averaging cycles. Figure \ref{fig:2} shows the value of \(H_N\left(x\right)\) for various \(N\) and \(x\). As one may expect, the accuracy function \(H_N\left(x\right)\)  is monotonically increasing for every \(N\), thus the estimated symbol selection accuracy is monotonically increasing in both \(n\) and \(\gamma\). We provide a proof of this property in Appendix \ref{app:B}.

\subsection{Accuracy estimation in practice}

In our model and derivation, we assumed that \(\Sigma\), \(\mu_0\) and \(\mu_1\) are known, yet in practice we usually use proxies in the form of their maximum-likelihood estimators,  \(\hat{\Sigma}\), \(\hat{\mu}_0\) and \(\hat{\mu}_1\). As those are dependent, however, their unbiasedness does not guarantee the unbiasedness of the accuracy estimation based on those estimates, and in fact the naïve estimation of the signal-to-noise ratio 
\(\hat{\gamma}=\sqrt{\left(\hat{\mu}_1-\hat{\mu}_0\right)^T\hat{\Sigma}^{-1}\left(\hat{\mu}_1-\hat{\mu_0}\right)}\)
 is indeed biased \cite{Lachenbruch1968EstimationAnalysis}. While exploring possible corrections for such estimation is beyond the scope of this article, we propose looking at this naïve estimate as a measure-of-quality of the BCI; we will demonstrate empirically that empirical SNR is a powerful predictor to the relation between the number of averaging cycles and symbol selection accuracy.

\section{Application to experimental data}
\label{section:Application}

We recorded the EEG signal of nine healthy individuals (2 female, aged 23-27 and 25 in average) during a standard P300 spelling task, each spelling 50 symbols (sessions lasted ~30 minutes). For more information regarding the experimental procedure, see Appendix \ref{app:C}. We then calculated the empirical symbol selection accuracy as a function of averaging cycle number using the following validation technique:

\begin{enumerate}
\item 	Train an LDA classifier from a random permutation of \(N_{train}=10\) symbols.
\item 	Evaluate the prediction on the remaining symbols.
\item   Repeat 1-2 for \(N_{reps}\). We chose \(N_{reps}=100\), which empirically ensures convergence.
\end{enumerate}

The reasoning behind using the above technique is that a traditional cross-validation fails to mimic the practical use of a P300 BCI, which typically relies on a small number of training samples. Additionally, this method ensures that different trials associated with the same symbol are never used in both train and test sets, which is important as P300 signal overlapping could lead to overfitting otherwise.

\begin{figure}
\centering
\includegraphics[width=0.8\textwidth]{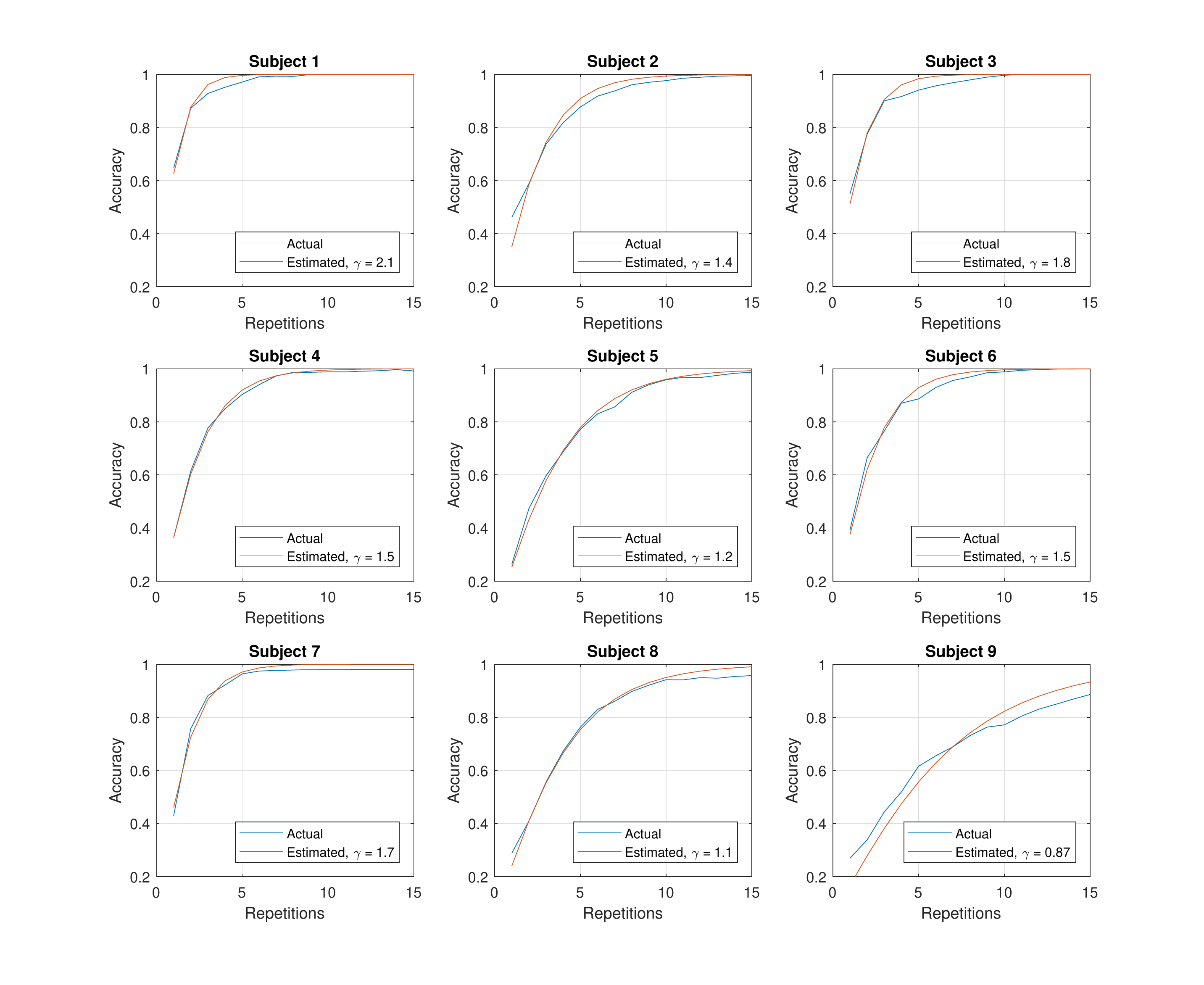}
\caption{\label{fig:3}The symbol selection accuracy vs. the number-of-repetitions curve, and the corresponding estimated accuracy fitted curve. Fit was selected to minimize the \(L^2\) distance between the actual and the fitted accuracy.}
\end{figure}

Figure \ref{fig:3} demonstrates the effectiveness of this estimation scheme for matching the accuracy-repetition curve: we calculated the empirical symbol selection accuracy for each possible number of repetitions, and then selected the SNR value \(\gamma\) to minimize the squared error of the prediction. The result is a curve that matches the accuracy for both small and large number of repetitions, for all subjects.

While the empirical SNR \(\hat{\gamma}=\sqrt{\left(\hat{\mu}_1-\hat{\mu}_0\right)^T\hat{\Sigma}^{-1}\left(\hat{\mu}_1-\hat{\mu_0}\right)}\) leads to a biased estimation of the accuracy, it seems that this bias is linear (Figure 4), and thus the empirical SNR can be used as a proxy for the accuracy. In Figure 5 we present a comparison between using the empirical SNR and using various other proxies for accuracy \cite{Farwell1988TalkingPotentials} – two versions of the peak-to-peak difference of the P300 signal and the area under the P300 curve – by looking at their relation to the accuracy (for a fixed number of repetitions), and demonstrate the effectiveness of the empirical SNR as a predictor of the accuracy.

\begin{figure}
\centering
\includegraphics[width=0.8\textwidth]{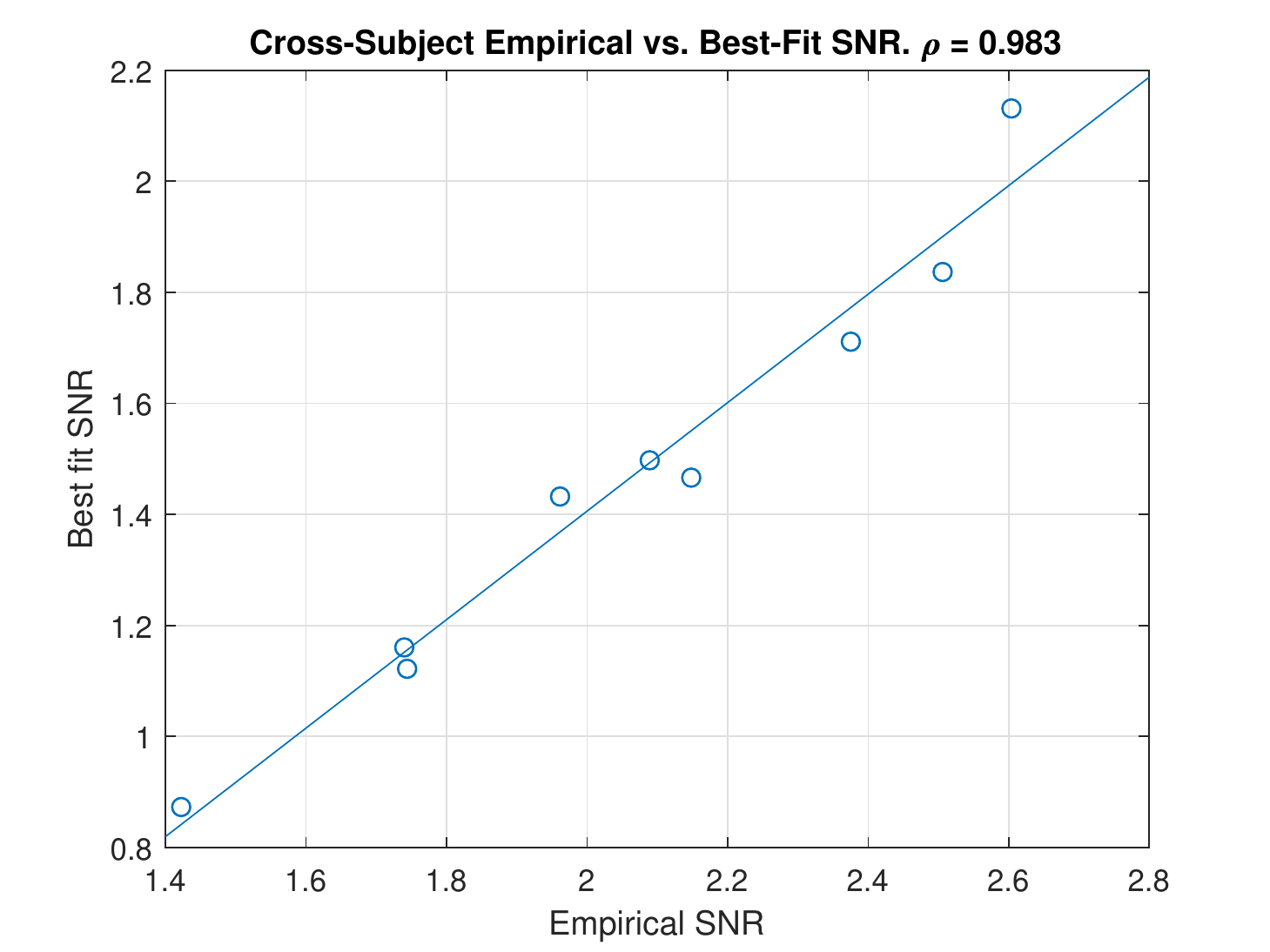}
\caption{\label{fig:4}Across-subject relation between the empirical SNR \(\hat{\gamma}\) and the value of \(\gamma\) that would best fit the accuracy-repetitions curve. We observe a linear relation (\(p=2\cdot 10^{-6}\)).}
\end{figure}

\begin{figure}
\centering
\includegraphics[width=0.8\textwidth]{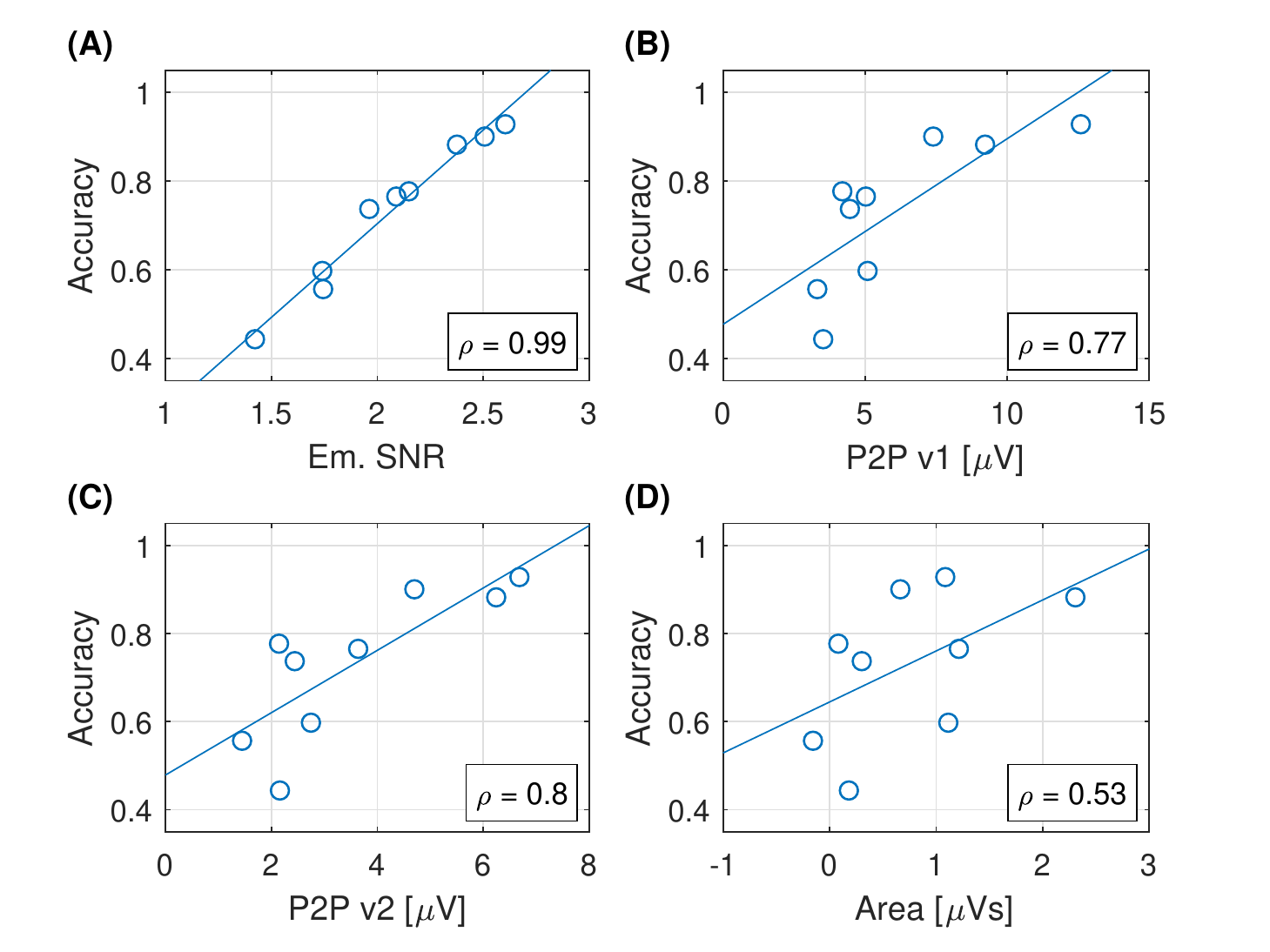}
\caption{\label{fig:5}Comparison of proposed proxies for the accuracy. Accuracy calculated for 3 repetitions (in order to avoid saturation), line shows a linear regression. Top-left: our proposed empirical SNR, \(\hat{\gamma}\). Top-right: version 1 of a peak-to-peak measure: \(\max{\left(\hat{\mu}_1-\hat{\mu}_0\right)}\). Bottom-left: version 2 of a peak-to-peak measure: \(\max{\left(\hat{\mu}_1\right)}-\max{\left(\hat{\mu}_0\right)}\). Bottom-right: the area under the P300 wave curve, \(\sum{\left(\hat{\mu}_1-\hat{\mu}_0\right)}\).}
\end{figure}

\begin{figure}
\centering
\includegraphics[width=0.8\textwidth]{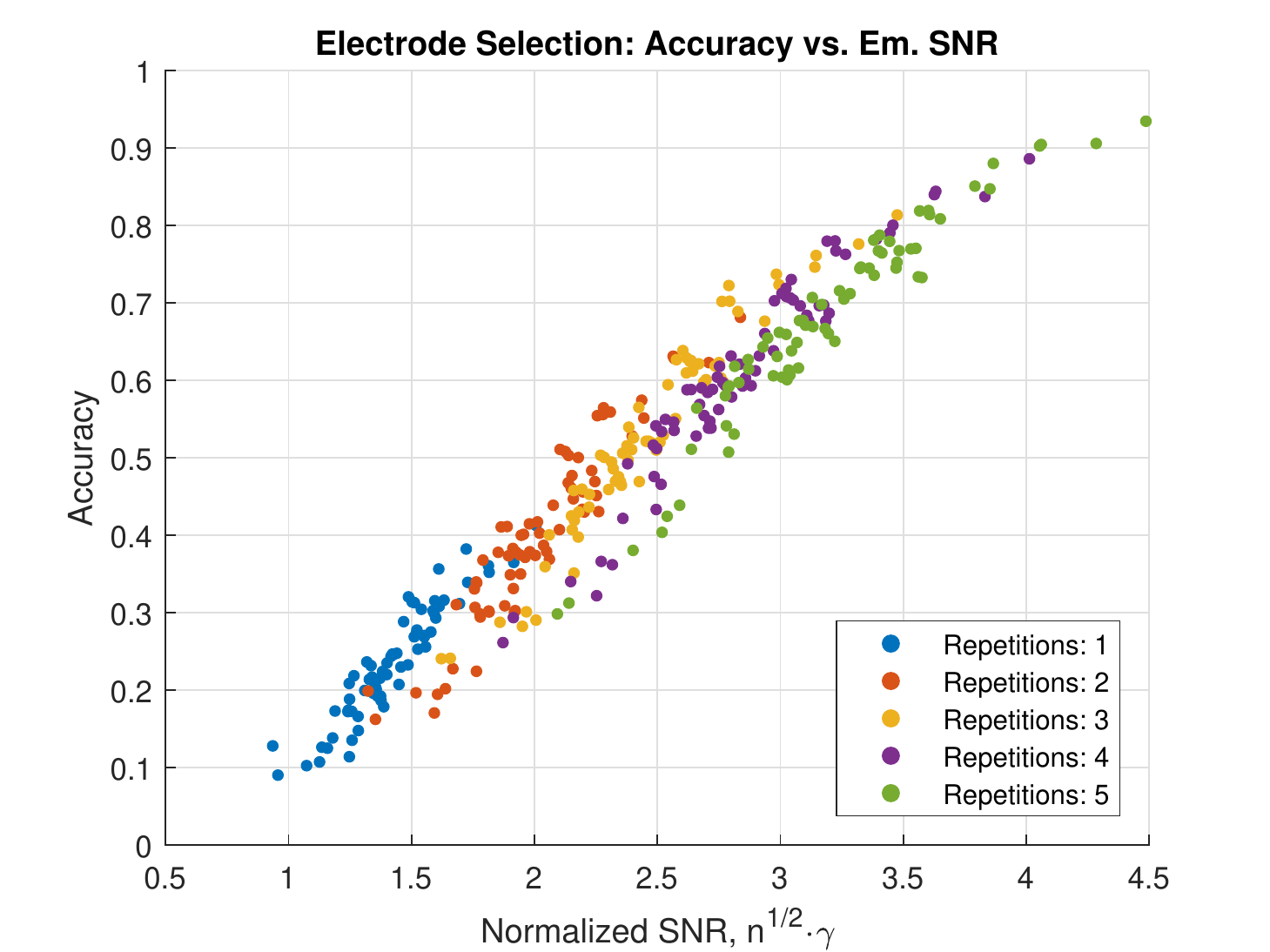}
\caption{\label{fig:6}Utilizing the empirical SNR for electrode subset selection. The SNR was multiplied by \(\sqrt{n}\) according to the number of repetitions n to allow for displaying multiple repetition values on a single graph.}
\end{figure}

Another task for which we can use the empirical SNR estimation is channel selection. Suppose we could sample only a limited number of electrodes in real-time due to some physical constraints, and want to decide which are most informative after a brief preparation session. Ideally, we would like to consider electrode selection as a hyperparameter, and train and validate models for all possible electrode subsets. There are, however, two difficulties with this approach:

\begin{enumerate}[label=\alph*.]
\item We might not have sufficient data to avoid overfitting, and would like to avoid investing a long time in collecting additional data for parameter selection;
\item Validation is a computationally demanding task, which requires multiple model training and testing epochs.
\end{enumerate}
In contrast, using the empirical SNR requires less data, not necessarily in a symbol-by-symbol form, and calculating the measure is much easier. We tested this approach for brute-force electrode selection for seven-out-of-eight electrodes, and the results are shown in Figure \ref{fig:6}. In our system, we observed a 400-fold speedup in calculating the empirical SNR compared to the validation accuracy.

\section{Discussion}
\label{section:Discussion}

Performance in P300-based BCIs exhibits high inter-subject variability. Whereas some subjects manage to achieve high accuracy with a few repetitions of the possible stimuli, others require many repetitions to achieve the same level of accuracy. In practice, this heterogeneity requires substantial fine tuning of parameters and extensive measurements from each subject to identify the optimal parameter set and evaluate performance.

In this study, we show that the signal-to-noise ratio of the P300 evoked potential, which can be estimated on a short time-scale using relatively few measurements, can serve as an accurate predictor for subject's performance. Our mathematical model assumes that the P300 response can be decomposed into a mean response component and a Gaussian noise component. Based on this model, we derived an expression for accuracy as a function of the number of repetitions. Surprisingly, this simple model provided highly accurate fits to the empirical data. Furthermore, SNR can be used to efficiently rank different parameter sets. For example, we demonstrated that it can be used to rank different electrode subsets. Because SNR is easily calculated, computation time is significantly shorter than in traditional cross-validation techniques. In addition, SNR is substantially more correlated with accuracy compared to other estimators, such as peak-to-peak of the P300 potential.

Clearly, the Gaussian model cannot fully capture the P300 response variability. However, the LDA scalar classification score is a linear combination of many measured EEG signals, each of which is in itself a superposition of many electric field sources. The central limit theorem suggests that it is reasonable to model the classification score as normally distributed, which is our key assumption. Another potential criticism of the model is the use of a linear classifier, a relatively weak classification technique. However, LDA is known to be the Bayes optimal classifier for classifying two Gaussian sources with the same covariance matrix. Thus, even if the sources are not precisely Gaussian, the LDA would still be close to optimal performance.

Even under our normality assumption, further refinements of the model are possible. First, we did not address the fact that the LDA's weight vector is in fact a random vector, and depends on the training set. Another issue is the fact that some measured signals overlap, violating our i.i.d. assumption. While both can be modeled, we decided to keep the model as simple as possible and leave this task for future work.

Another open issue is connecting the empirical SNR to the best-fit SNR. We pointed out a linear empirical relation between these quantities. It might be possible to explain this linear relation through further mathematical exploration of our model. Furthermore, while this empirical relation could be used to correct for SNR estimations for the specific acquisition parameters in our system, future work could generalize this relation to a wider range of systems and acquisition parameters.

Finally, we believe that the model could have a wide range of applications beyond those described here. In particular, our model connects the basic properties of the P300 evoked potential to the performance of the BCI. For example, when considering the use of a P300 BCI for an individual, such as an ALS patient, it would suffice to perform a basic oddball experiment to characterize the P300 SNR before adapting an expensive BCI system. Another relevant scenario is that of testing a novel P300 paradigm that might improve performance. For example, testing the effect of the symbol matrix size on performance is tedious in the traditional approach and requires many BCI spelling experiments for different matrix sizes \cite{Sellers2006APerformance}. Using the approach proposed here, an off-the-shelf oddball paradigm experiment could be conducted, measuring the SNR for different numbers of stimuli. Finding the optimal matrix size could then be achieved by quantitively estimating how the SNR boost and the number of elements in the symbol matrix affect the bit rate. In general, wide usage of SNR estimations would allow a faster paradigm evaluation, and ultimately faster and less expensive BCI development.

\section*{Disclosure statement}

The authors report no conflict of interest.

\section*{Acknowledgments}

We would like to thank Tomer Fekete and Miriam Nagel for their helpful comments. This research was supported in part by the Helmsley Charitable Trust through the Agricultural, Biological and Cognitive Robotics Initiative and by the Marcus Endowment Fund both at Ben-Gurion University of the Negev.


\bibliographystyle{tfnlm}
\bibliography{references}

\appendix
\section{Derivation of the accuracy function}
\label{app:A}

The basic premises of our model are:

\begin{enumerate}
\item 	The base P300 signal is generated deterministically, so it should have been detected at the EEG as \(\mu_1\) whenever it is generated. When no P300 signal is generated, one would detect a baseline signal of \(\mu_0\).
\item	The actual sampled signal is the P300 wave plus some noise, either 
\(x=\mu_1+z\) or \(x=\mu_0+z\). The noise is distributed normally with mean zero and covariance matrix \(\Sigma\). All the noise vectors are sampled i.i.d, and are independent of the presence of the P300 signal.
\item	The detection mechanism is as described in section 1.1, with an LDA classifier. Although the weight vector of the classifier is determined by a contingent training set and is therefore random, we assume for the sake of simplicity that it is indeed the Bayesian optimum of \(w=\Sigma^{-1}\cdot\left(\mu_1-\mu_0\right)\).
\end{enumerate}

Following these assumptions, the sampled signal corresponding to the i'th stimulus is

\[x_i=\mu_{y\left(i\right)}+n\sim N\left(\mu_{y\left(i\right)}, \Sigma\right)\]
with \(y\left(i\right)=1\) if the \(i\)'th stimulus triggered a P300 response and \(y\left(i\right)=0\) otherwise. The averaged signal for the \(i\)'th stimulus after \(n\) averaging cycles is then distributed as

\[\bar{x}_i\sim N\left(\mu_{y\left(i\right)}, \frac{\Sigma}{n}\right)\]

Assumption (3) implies that the classification score  \(\bar{s}_i=w^T\bar{x}_i\) is also normally distributed,

\[\bar{s}_i\sim N\left(m_{y\left(i\right)}, \sigma^2_n\right)\]
with mean \(m_{y\left(i\right)}=\left(\mu_1-\mu_0\right)^T\Sigma^{-1}\mu_{y\left(i\right)}\) and variance \(\sigma^2_n=\frac{1}{n}\left(\mu_1-\mu_0\right)^T\Sigma^{-1}\left(\mu_1-\mu_0\right)\stackrel{\text{def}}{=}\frac{1}{n}\gamma^2\). Note that by our definition of the signal-to-noise ratio we have that \(\frac{m_1-m_0}{\sigma_n}=\sqrt{n}\gamma\).

The structure of the detection mechanism implies that the row and the column are independently selected, thus symbol selection accuracy is the product of the row and the column selection accuracies. We will discuss the row selection accuracy \(Acc_{row}\) for \(N_{row}\) rows, and then infer the column selection accuracy \(Acc_{col}\) and the symbol selection accuracy, \(Acc=Acc_{row}\cdot Acc_{col}\). 

Without loss of generality, we can consider row \(1\) to be the target row, and rows \(2, \dots ,N_{row}\) to be the non-target rows. Therefore, \(\bar{s}_1\sim N\left(m_1, \sigma^2_n\right)\) and \(\bar{s}_i\sim N\left(m_0, \sigma^2_n\right)\) for \(i=2, \dots ,N_{row}\). Defining 
\(\bar{s}_0^{max}=\max\limits_{i=2, \dots ,N_{row}}\bar{s}_i\)
 we obtain that the row selection accuracy is
\[Acc_{row}=\Pr\left(\bar{s}_1>\max\limits_{i=2, \dots ,N_{row}}\bar{s}_i\right)=\intop_{-\infty}^{\infty}f_{\bar{s}_1}\left(x\right)F_{\bar{s}_0^{max}}\left(x\right)dx\]
with \(f_{\bar{s}_1}\left(x\right)\) and \(F_{\bar{s}_0^{max}}\left(x\right)\) being the PDF of \(\bar{s}_1\) and the CDF of \(\bar{s}_0^{max}\), respectively. Since the scores \(\bar{s}_i\) for \(i=2, \dots ,N_{row}\) are i.i.d \(N\left(m_0, \sigma^2_n\right)\), we have
\[F_{\bar{s}_0^{max}}\left(x\right)=\prod_{i=2, \dots ,N_{row}} F_{\bar{s}_i}\left(x\right)=\Phi^{N_{row}-1}\left(\frac{x-m_0}{\sigma_n}\right)\]
and
\[\begin{split}
Acc_{row}&=\intop_{-\infty}^{\infty}\frac{1}{\sigma_n}\phi\left(\frac{x-m_0}{\sigma_n}\right)\Phi^{N_{row}-1}\left(\frac{x-m_0}{\sigma_n}\right)dx=\\
&=\intop_{-\infty}^{\infty}\phi\left(z-\frac{m_1-m_0}{\sigma_n}\right)\Phi^{N_{row}-1}\left(z\right)dx=\\
&=\intop_{-\infty}^{\infty}\phi\left(z-\sqrt{n}\cdot\gamma\right)\Phi^{N_{row}-1}\left(z\right)dx\stackrel{\text{def}}{=}
H_{N_{row}}\left(\sqrt{n}\cdot\gamma\right)
\end{split}\]
with the accuracy function \(H_N \left(x\right)\) defined as
\[H_N\left(x\right)
\stackrel{\text{def}}{=}
\intop_{-\infty}^{\infty}\phi\left(z-x\right)\Phi^{N-1}\left(z\right)dx
\]
where \(\phi\) and \(\Phi\) are the PDF and the CDF of a standard normal distribution, respectively. As row and column are selected independently, the estimated symbol selection accuracy will be
\[\begin{split}
Acc &=H_{N_{row}}\left(\sqrt{n}\cdot\gamma\right)\cdot H_{N_{col}}\left(\sqrt{n}\cdot\gamma\right)= \\
= &\left(\intop_{-\infty}^\infty \phi\left(z-\sqrt{n}\cdot\gamma\right)\cdot\Phi^{N_{row}-1}\left(z\right) dz\right)\cdot \\
\cdot &\left(\intop_{-\infty}^\infty \phi\left(z-\sqrt{n}\cdot\gamma\right)\cdot\Phi^{N_{col}-1}\left(z\right) dz\right)
\end{split}\]

\section{Monotonicity of the accuracy function}
\label{app:B}

Since the argument of the accuracy function in our expression is the effective SNR of the detection, we would expect \(H_N \left(x\right)\) to be monotonically increasing in \(x\). This is indeed true, as
\[\begin{split}
\frac{d H_N\left(x\right)}{d x}&=\intop_{-\infty}^{\infty}-\phi'\left(z-x\right)\cdot\Phi^{N-1}\left(z\right) dz=
\intop_{-\infty}^{\infty}-\phi'\left(y\right)\cdot\Phi^{N-1}\left(y+x\right) dy=\\
&=\intop_{-\infty}^{0}-\phi'\left(y\right)\cdot\Phi^{N-1}\left(y+x\right) dy+
\intop_{0}^{\infty}-\phi'\left(y\right)\cdot\Phi^{N-1}\left(y+x\right) dy=\\
&=\intop_{0}^{\infty}-\phi'\left(-y\right)\cdot\Phi^{N-1}\left(-y+x\right) dy+
\intop_{0}^{\infty}-\phi'\left(y\right)\cdot\Phi^{N-1}\left(y+x\right) dy=\\
&=\intop_{0}^{\infty}-\phi'\left(-y\right)\cdot\left(\Phi^{N-1}\left(x+y\right)-\Phi^{N-1}\left(x-y\right)\right) dy
\end{split}\]
where we used the fact that \(\phi\) is even. Next, we should note that \(\Phi\) is monotonically increasing thus \(\Phi^{N-1}\left(x+y\right)-\Phi^{N-1}\left(x-y\right)>0\), and that for positive \(y\) we get \(\phi'\left(y\right)<0\); so indeed \(\frac{d H_N\left(x\right)}{d x}>0\).

\section{Experimental procedure}
\label{app:C}

The EEG system that we employed consisted of a gTec's g.HIamp amplifier and g.LADYbird active electrodes. The impedance of the electrodes was kept under \(30\,k\Omega\) during the entire experiment. We used a \(6\times 6\) character matrix for a total of 36 different letters, numbers and symbols, utilizing the standard row-column paradigm. The flash duration was \(62.5\,ms\) and the ISI was \(125\,ms\). Because there were 12 stimuli in total (6 rows and 6 columns), the total time of one flashing cycle was \(1.5\,s\). After each spelled character, there was an \(8\,s\) pause in which the next target character was intensified. As suggested by previous works \cite{KrusienskiASpeller}, the EEG signal was measured using the following 8 electrodes: Fz, Cz, P3, Pz, P4, Po7, Oz, Po8 (all according to the international 10-20 system), with the reference at the left ear. The sampling frequency was \(256\,Hz\), which was immediately downsampled (by 4:1 averaging) to \(64\,Hz\). A bandpass filter of \(0.5\)-\(30\,Hz\) and a notch filter at the network frequency (\(50\,Hz\)) were applied. For each flashing sequence, the time segment was considered as \(600\,ms\) from the beginning of the flash (at \(64\,Hz\), this is 39 samples. Note that since the ISI was smaller than the segment length, the segments overlapped), and the samples from the different electrodes were concatenated into one high-dimensional vector (in our case, 312 elements).

The experiment was performed on 9 participants (2 female, aged 23-27 and 25 in average, all undergrad students at BGU) with no previous P300 BCI experience and with normal or corrected to normal vision. Each was asked to spell 10 five-letter words (50 symbols in total), spelling the characters in each word sequentially and possibly taking a break between the words. Each symbol consisted of 15 repetitions for each row or column flash, or a total of 180 flashes per symbol, lasting \(22.5\) seconds. Before the acquisition session started, participants practiced counting the flashes and familiarized themselves with the physical setup for about one minute (this part was not recorded or processed). The subjects received feedback (in terms of estimated character) after every character, based on a classifier that was trained after the first five-letter word (which was not used in the offline analysis).

\end{document}